\documentclass[singlespacing]{elsart}

\usepackage{epsfig}

\usepackage{amssymb}

\begin{document}

\begin{frontmatter}



\title{Charged particle densities 
from Au+Au collisions at  $\sqrt{s_{\small{NN}}}$=130 GeV}


\author{}
\author[nbi]{I.~G.~Bearden},
\author[bnl]{D.~Beavis},
\author[bucharest]{C.~Besliu}, 
\author[newyork]{Y.~Blyakhman},
\author[krakow]{J.~Brzychczyk},
\author[newyork]{B.~Budick},
\author[nbi]{H.~B{\o}ggild}, 
\author[bnl]{C.~Chasman},
\author[nbi]{C.~H.~Christensen}, 
\author[nbi]{P.~Christiansen}, 
\author[kraknuc]{J.~Cibor}, 
\author[bnl]{R.~Debbe}, 
\author[nbi]{J.~J.~Gaardh{\o}je}, 
\author[krakow]{K.~Grotowski}, 
\author[texas]{K.~Hagel}, 
\author[nbi]{O.~Hansen}, 
\author[nbi]{A.~Holm},
\author[oslo]{A.~K.~Holme}, 
\author[kansas]{H.~Ito}, 
\author[nbi]{  E.~Jakobsen}, 
\author[bucharest]{A.~Jipa}, 
\author[bergen]{J.~I.~J{\o}rdre}, 
\author[ires]{F.~Jundt},
\author[nbi]{  C.~E.~J{\o}rgensen}, 
\author[texas]{T.~Keutgen}, 
\author[baltimore]{E.~J.~Kim}, 
\author[krakow]{T.~Kozik}, 
\author[oslo]{T.~M.~Larsen}, 
\author[bnl]{J.~H.~Lee}, 
\author[baltimore]{Y.~K.~Lee},
\author[oslo]{G.~L{\o}vh{\o}iden}, 
\author[krakow]{Z.~Majka}, 
\author[texas]{A.~Makeev},
\author[bnl]{E.~McBreen}, 
\author[texas]{M.~Murray}, 
\author[texas]{J.~Natowitz}, 
\author[nbi]{B.~S.~Nielsen}, 
\author[bnl]{K.~Olchanski},
\author[bnl]{J.~Olness}, 
\author[nbi]{D.~Ouerdane}, 
\author[krakow]{R.~P\l aneta}, 
\author[ires]{F.~Rami}, 
\author[bergen]{D.~R{\"o}hrich}, 
\author[oslo]{B.~H.~Samset},
\author[kansas]{S.~J.~Sanders\corauthref{cor1}}, 
\author[bnl]{R.~A.~Sheetz}, 
\author[krakow]{Z.~Sosin}, 
\author[nbi]{P.~Staszel},
\author[bergen]{T.~F.~Thorsteinsen\thanksref{deceased}}, 
\author[oslo]{T.~S.~Tveter}, 
\author[bnl]{F.~Videb{\ae}k}, 
\author[texas]{R.~Wada}, 
\author[krakow]{A.~Wieloch}, 
\author[bucharest]{I.~S.~Zgura}, 
\author{(BRAHMS Collaboration )}

\address[bnl]{Brookhaven National Laboratory, Upton,New York 11973}
\address[ires]{Institut de Recherches Subatomiques and Universit{\'e} 
Louis Pasteur, Strasbourg, France}
\address[kraknuc]{Institute of Nuclear Physics, Krakow, Poland}
\address[krakow]{Jagiellonian University, Krakow, Poland}
\address[baltimore]{Johns Hopkins University, Baltimore,Maryland 21218}
\address[newyork]{New York University, New York, New York 10003}
\address[nbi]{Niels Bohr Institute, University of Copenhagen, Denmark}
\address[texas]{Texas A$\&$M University, College Station,Texas 77843}
\address[bergen]{Department of Physics, University of Bergen, Bergen,Norway}
\address[bucharest]{University of Bucharest,Romania}
\address[kansas]{Department of Physics and Astronomy, University of Kansas, Lawrence, Kansas 66045}
\address[oslo]{Department of Physics, University of Oslo, Oslo,Norway}
\thanks[deceased]{deceased}
\corauth[cor1]{Corresponding author, e-mail: SSanders@ku.edu}
\begin{abstract}
We present charged particle densities as a function of pseudorapidity and
collision centrality for the $^{197}$Au+$^{197}$Au reaction
at $\sqrt{s_{NN}}$=130 GeV.  An integral charged particle multiplicity of
3860$\pm$300 is found for the 5\% most central events within the 
pseudorapidity range $-4.7\le \eta \le 4.7$.  At mid-rapidity an
enhancement in the particle yields per 
participant nucleon pair is observed for central events.
Near to the beam rapidity, 
a scaling of the particle yields consistent with
the ``limiting fragmentation'' picture is observed.  
Our results are compared to other recent 
experimental and theoretical discussions
of charged particle densities in ultra-relativistic heavy-ion collisions.
\end{abstract}

\begin{keyword}

\PACS 25.75.Dw
\end{keyword}
\end{frontmatter}


Multiplicity distributions of emitted 
charged particles  provide a fundamental measure
of the ultra-relativistic collisions 
now accessible experimentally using 
the Relativistic Heavy-Ion Collider (RHIC).  
The particle densities are sensitive to the
relative contribution of ``soft'' processes, involving the 
longer length scales associated with non-perturbative
QCD mechanisms,  and ``hard'',  
partonic processes~\cite{wang00,eskola00}. 
The total number of charged particles and the 
angular dependence of the charged particle  distribution 
is expected to depend  markedly on the amount of hadronic rescattering, 
the degree of chemical  and thermal equilibration, and the role of 
subnucleonic processes. 
 
In this Letter we present charged particle distributions
in $dN_{ch}/d\eta$ with $-4.7 \le \eta \le 4.7$ for $^{197}$Au~+~$^{197}$Au
collisions at $\sqrt{s_{\small{NN}}}$=130 GeV.  The pseudorapidity
variable $\eta$ can be expressed in terms of the scattering angle
$\theta$, with
$\eta=-{\rm ln}[{\rm tan}(\theta/2)]$.
The data were measured at the RHIC facility 
using complementary subsystems of the  BRAHMS experiment.
We have also investigated the systematics of 
the pseudorapidity  distributions as a function of collision  
centrality. The collision centrality can be related 
to the number of participant nucleons in the reaction, thus
allowing for comparison of different nuclear systems based on
simple nucleon-nucleon superposition models~\cite{bialas76}, 
as has been done, for example, at lower SPS energies for the 
Pb+Pb system~\cite{deines00,aggarwal01}. Our results
complement those recently reported by the PHOBOS~\cite{back00,back01} 
and PHENIX~\cite{adcox01} collaborations by establishing global
$dN_{ch}/d\eta$ systematics through a very different 
experimental arrangement than used in those experiments.

\begin{figure}[htp]
\epsfig{file=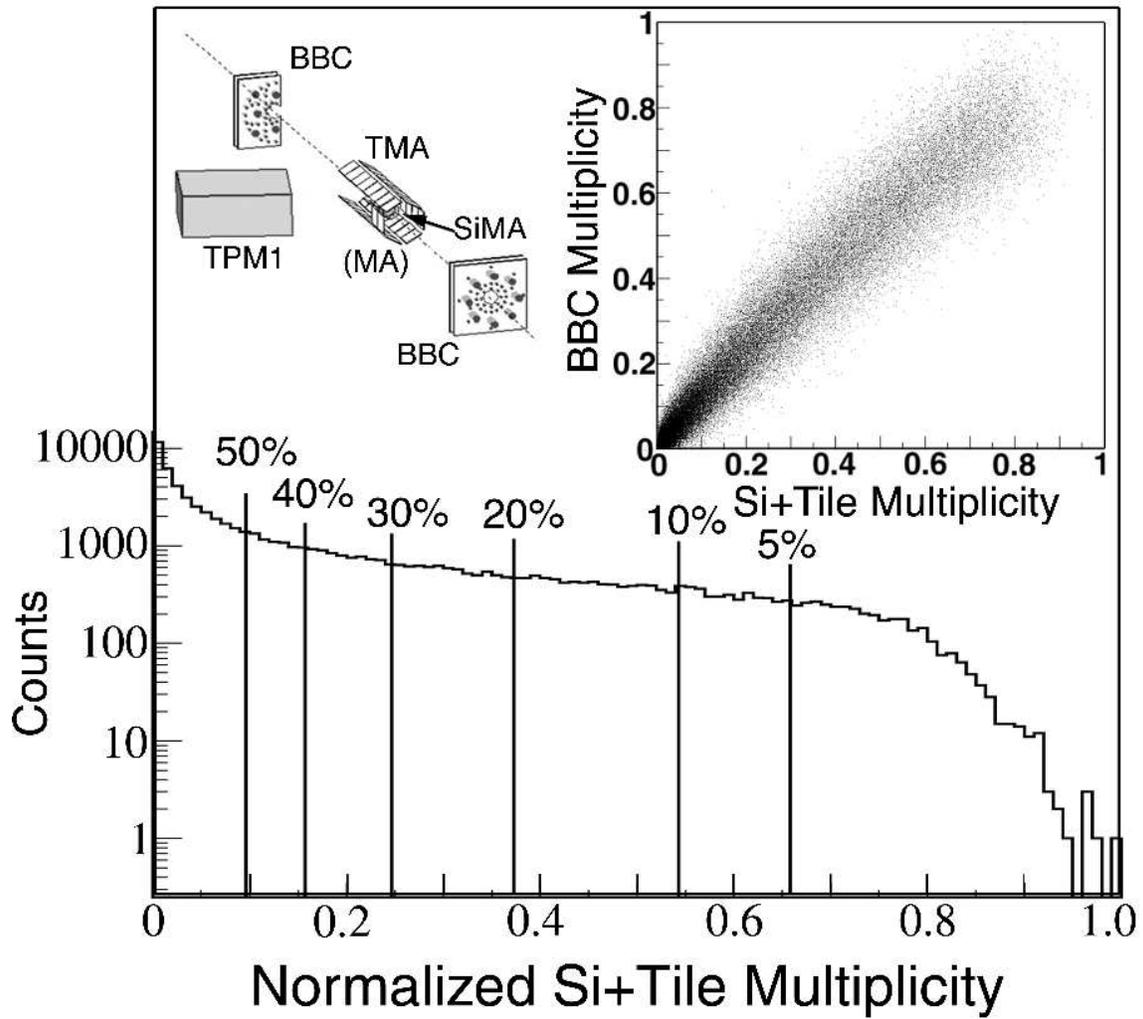,width=15.0cm}
\caption{
Normalized MA array multiplicity distribution, as discussed in the text.
The scatter plot insert shows the correlation of the
normalized BBC and MA multiplicities.   
A schematic drawing of the global detectors
used in the measurement, excluding the ZDCs, is also
shown.
}
\label{mult}
\end{figure}

The BRAHMS detector system~\cite{BRAHMSNIM} consists of forward- and 
mid-rapidity
magnetic spectrometers, 
which allow for the determination of charged particle 
properties over a wide rapidity and momentum range, and
a number of global detectors employed to characterize the 
general features of the reaction, such as the overall charged 
particle multiplicity and the flux of spectator neutrons at
small angles.  
The current analysis is based primarily on the global detectors 
(see Fig.~\ref{mult} insert),
including the Multiplicity Array (MA), the Beam-Beam counter arrays
(BBC), and the Zero-Degree Calorimeters (ZDC)~\cite{adler00}.   
Also, the front time-projection chamber (TPM1) of
the mid-rapidity spectrometer arm, which is an integral part of this
spectrometer, provides independent confirmation of the
particle densities at selected angles. 

The MA is used to determine  charged
particle densities in the pseudorapidity range of
$-2.2\le\eta\le2.2$ for collisions at the nominal vertex. 
This pseudorapidity coverage is increased to $-3.0\le\eta\le3.0$ 
by using an extended range of vertex locations. For the present data, 
the collision vertex distribution was
found to be roughly Gaussian in shape with $\sigma$=70~cm.  
The MA consists of two layers of
detectors arranged as  six-sided coaxial barrels about the beam axis.
The inner barrel (SiMA) is comprised of 
25, 4~cm$\times$6~cm$\times$300~$\mu$m Si 
strip detectors  located 5.3~cm
from the nominal beam axis, with each detector 
segmented into 7 strips with a 0.86~cm pitch. 
For the SiMA, three sides were populated with six wafers each, 
covering 42 segments in
pseudorapidity, one side was populated with five wafers, 
and the last two sides were populated with one wafer, each. 
The outer barrel (TMA)  
consists of 38, 12~cm$\times$12~cm$\times$0.5~cm
plastic scintillator ``tile'' detectors~\cite{aota95} located 13.9~cm from 
the beam axis. The scintillator tile array had four rows fully populated
with 8 detectors, each, one row had 4 detectors, and the last row had 2
detectors.    

The single-particle response of the SiMA and TMA elements was calibrated
by selecting peripheral collision events
where energy peaks corresponding to the passage of single 
particles, mostly pions,  through the detectors could be compared
to the calculated energies for these peaks based on GEANT 
simulations~\cite{geant} and assuming nominal detector thicknesses. 
The uncertainty in the single-particle response
is estimated as $\approx$5\% 
based on the scatter of measurements
for a given detector element for different vertex locations and
based on the uncertainty in the GEANT simulations.    

For more central collisions at RHIC, the modest segmentation 
of the MA results in multiple particles passing
through individual detector elements. The number of particles passing
through each element is determined by dividing the total energy
observed in  that element by the corresponding average energy loss for a single
particle, primarily either a pion, kaon, or proton,  
as determined by GEANT simulations. Near to mid-rapidity and for central
collisions an average of  10 particles pass through the Si elements
and 61 particles traverse the tile elements.  
Two different event simulators, 
HIJING~\cite{wang91} and Fritiof~\cite{pi92},  
were used to check the sensitivity of the analysis
procedure to the assumed primary particle distributions and the corresponding 
momentum distributions. 
The deduced particle
multiplicities using the different codes agree to better than 3\%. This
uncertainty is folded into the quoted overall systematic uncertainties. 
The particle yields are corrected for background contributions
using the GEANT simulations of the array response.  These corrections are
position dependent and range from 20\% to 40\% for the TMA and from
6\% to 25\% for the SiMA, increasing with increasing $|\eta|$.

The BBC Arrays consist of two sets of 
Cherenkov UV transmitting plastic radiators 
coupled to photomultiplier tubes. They are positioned around the 
beam pipe on either side of the nominal interaction point at a 
distance of 2.15~m. One array consists of 8 ``large''
detectors with  51~mm 
diameter phototubes and 36 ``small'' detectors with 19~mm diameter 
phototubes, 
arranged symmetrically 
around the beam pipe. The other array is asymmetric to allow the 
movement of the forward spectrometer to small angles and consists 
of 5 large detectors and 30 small detectors. The BBC elements 
have an intrinsic time resolution of 65~ps allowing for the determination 
of the position of the interaction point with a precision of 
$\approx$ 1.6~cm. 
Charged particle
multiplicities in the pseudorapidity range $2.1\le |\eta| \le 4.7$ 
are deduced from the number of particles hitting each tube, as found
by dividing the measured ADC signal by that corresponding to a
single particle hitting the detector.  Corrections
for background events were based on GEANT simulations,
with the correction factors ranging from 37\% to  50\% of the measured yield,
depending on vertex location.

The two ZDCs, positioned on either side of the nominal interaction 
point at a distance of 18~m, were used to establish a minimum bias trigger 
for the experiment.  The ZDCs,
which consist of alternating 
layers of tungsten and fiber-ribbon plastic scintillator,  
measure the energy 
deposited by spectator neutrons that are emitted at small angles
with respect to the beam direction, with $|\theta|< 2$~mrad~\cite{adler00}. 
The time difference between the 
two ZDC signals can be used to determine the interaction vertex 
with a precision of $\approx$~3.6~cm.

In addition to the global detector measurements,
charged particle multiplicities were also obtained at 
40$^\circ$, 60$^\circ$, and
90$^\circ$, corresponding to $\eta$=1.1, 0.55, and 0.0, respectively,
using the TPM1 counter.  
This detector, which is operated in a field-free region, 
has an active volume of (w,h,d)=(37.5~cm, 21.0~cm, 36.0~cm)
with its volume center located 94.5~cm from the  nominal vertex. 
For TPM1, particle multiplicities are deduced by 
counting reconstructed tracks arising from collision
vertices within 15~cm of the nominal vertex position. 
The
acceptance in $\eta$ for this counter, accounting for both the geometric
acceptance and the range of vertex locations,  
varies from $\Delta\eta$=0.5 at 
90$^\circ$ to $\Delta\eta$=0.60 at 40$^\circ$. 
Through the track reconstruction,
the interaction vertex can be located to better than 0.5~cm.   
A correction for the background of secondary particles was found by
projecting the vertical position of tracks in TPM1 back to the beamline.
The resulting spectrum was fitted with the sum of three Gaussian 
distributions; one for
the peak of primary particles and summed background contributions from weak
decays or secondary interactions, and delta electrons from the beampipe,
respectively. The ratio of the integral of the signal peak to the two
background distributions was taken as the fraction of primary particles. 
Typical background-to-total ratios of $\approx$5\% were found at all
angles.

The dominant systematic uncertainties  for
the TPM1 measurements arise from the background
subtraction and a tracking efficiency
correction. 
The efficiency correction is obtained by inserting simulated
single tracks into actual data events and analyzing 
the reconstruction probability using the same
track reconstruction code employed for the experimental results. 
The tracking
efficiency in the TPM1 counter ranges from almost 
unity for low density, peripheral scattering events, 
to $\approx$0.9 for central events at 90$^\circ$ and $\approx$0.85
for central events at 40$^\circ$.
In contrast, the dominant systematic uncertainty near mid-rapidity for
the SiMA and TMA arrays reflect the single-particle
response calibrations obtained for the
constituent  detectors of these arrays. 
At larger pseudorapidity values, the uncertainty in the background
subtraction also becomes significant. 
In addition, the relatively thick 
scintillator elements of the TMA array are expected to be somewhat
sensitive to
beam related background that does not come from the collision vertex.   
These effects lead to the somewhat larger uncertainty quoted for the TMA as 
compared to the SiMA.
The BBC systematic uncertainty is dominated by the background 
correction.  Overall, the systematic uncertainties for the 
different detector
systems are: SiMA, $\pm$8\% for $\eta$$<$1.5, increasing to
$\pm$10\% for $\eta\ge$2.5; TMA, $\pm$12\% for $\eta<$1.5 increasing to
$\pm$15\% for $\eta\ge$2.0; BBC, $\pm$10\%; TPM1, +9\%/-7\% for the 
most central events and $\pm$6\% for the most peripheral. 

The centrality selection for the experiment was obtained by developing
a minimum-biased multiplicity distribution using the MA, 
reflecting all events for which 
there is a nuclear interaction, and assuming 
that a cut on the total multiplicity translates to a cut 
on collision centrality.    
Fig.~\ref{mult} shows the 
multiplicity distribution of charged particles 
established by the MA, normalized to the maximum observed multiplicity. 
The independent multiplicity measurements of the
SiMA and TMA detectors are summed for this figure, after
the SiMA multiplicity was rescaled to account for the difference in
the geometric coverage of the two arrays.  Where possible,
the location of the 
interaction vertex was determined using TPM1.
Otherwise, the vertex determined from the BBC or ZDC data was adopted. The charged 
particle densities reported in this paper are obtained almost exclusively using the
TPM1 and BBC vertices and are relatively insensitive to the specific choice.  
The multiplicity distribution requires that
a coincidence occurs between the two ZDC detectors, an interaction vertex
be located within 30 cm of the array center (as obtained by one of the three
possible vertex measurements), and that a TMA multiplicity of at least four
be observed.  The TMA
multiplicity cut eliminates events corresponding to 
beam-gas interactions and for cases where tails of pedestals for individual
detector elements add to simulate real collision events. 
Based on HIJING simulations, where the shape of the multiplicity distribution
for events with fewer than 200 tile hits is used to estimate the 
missing yield for events with fewer than four tile hits,  
it is estimated that the TMA array
observes cleanly
95\% of the total nuclear cross sections. This value is used to correct for 
the missing cross section where the tile multiplicity is less than
four when determining the final multiplicities.    

Centrality selection in the BBC analysis was done 
using the multiplicity distribution
obtained by summing the hits in the BBC arrays.  This allows for the use of 
interaction vertices within 120 cm of the MA center, well outside the
range for which the MA multiplicities could be reliably determined. The
insert in Fig.~\ref{mult} shows the strong correlation of the
BBC and MA multiplicities, normalized to their maximum values. 
In the pseudorapidity range of 3.0$\le\eta\le$4.2,
where it was possible to analyze the BBC data using both
centrality selections, the two analyses give identical results
to within 1\%.  Still, the lower segmentation of the BBC arrays leads to
a less precise multiplicity measurements as compared to the MA.

\begin{figure}
\epsfig{file=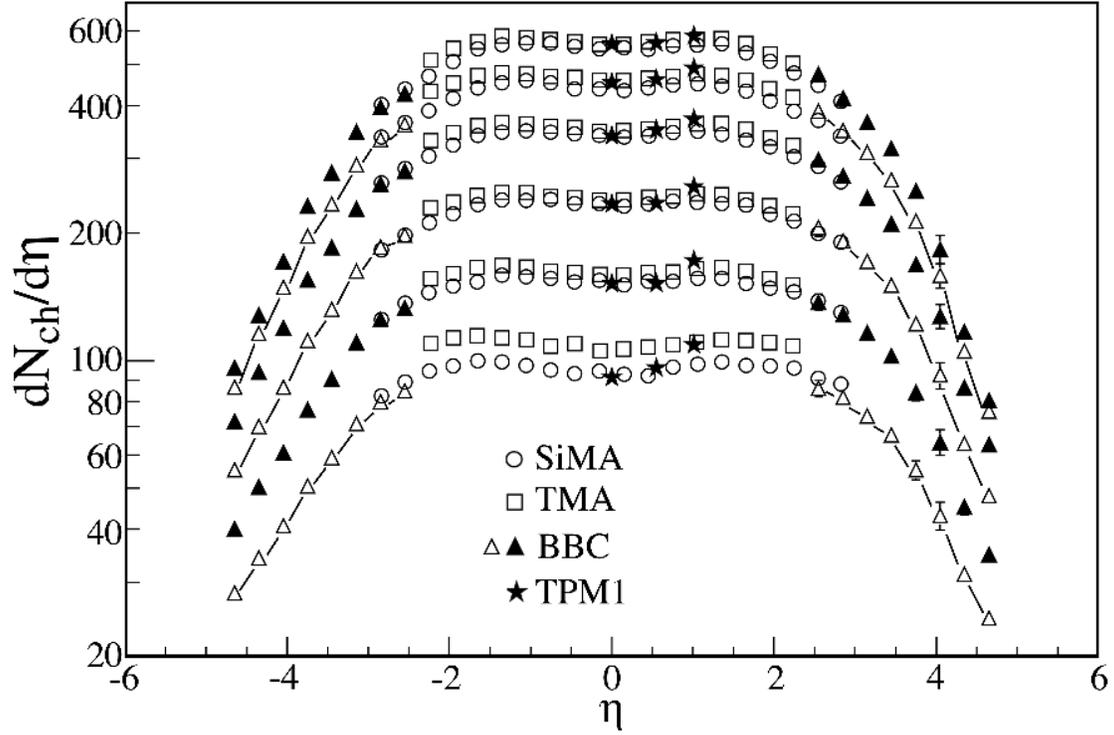,width=15.0cm}
\caption{
Distributions of  $dN_{ch}/d\eta$ for centrality ranges of,
top to bottom, 0-5\%, 5-10\%, 10-20\%, 20-30\%, 30-40\%, and
40-50\%. 
Statistical uncertainties are shown where larger than the symbol size.
The connecting lines and alternating open and closed symbols for the
BBC data are to help distinguish points associated with 
different centrality ranges. 
 }
\label{dndeta}  
\end{figure}

Figure~\ref{dndeta} 
presents the measured dN$_{ch}$/d$\eta$ distributions for a number of
centrality cuts.  
The points associated with positive pseudorapidity values correspond
to detectors located on the forward spectrometer side of the 
BRAHMS apparatus. Within
the quoted systematic uncertainties, good
agreement is found among the  measurements using different detector systems. 
There is, however, an offset 
observed for the TMA data which may reflect non-collision related background 
events. The other detectors are relatively insensitive to such a background
mechanism. 

Particle densities at selected pseudorapidities are tabulated in Table 1.
This table also presents the integrated charge-particle multiplicities
in the range $-4.7\le\eta\le4.7$.  For central
events at mid-rapidity, the observed particle densities are about 1.8 times 
greater than observed for central events in Pb+Pb scattering at 
$\sqrt{s_{\small{NN}}}$=17.2 GeV~\cite{deines00}.

The PHOBOS~\cite{back00} and PHENIX~\cite{adcox01} experiments have 
recently reported values for the charged particle densities at mid-rapidity
corresponding to the 6\% most central events. PHOBOS obtains 
$dN_{ch}/d\eta_{|\eta=0}=555\pm12(stat)\pm35(syst)$ while
PHENIX finds 
$dN_{ch}/d\eta_{|\eta=0}=609\pm1(stat)\pm37(syst)$.  For the
same centrality range, our result is
$dN_{ch}/d\eta_{|\eta=0}=549\pm1(stat)\pm35(syst)$. Within the systematic
uncertainties the three experiments are in agreement.

\begin{figure}
\epsfig{file=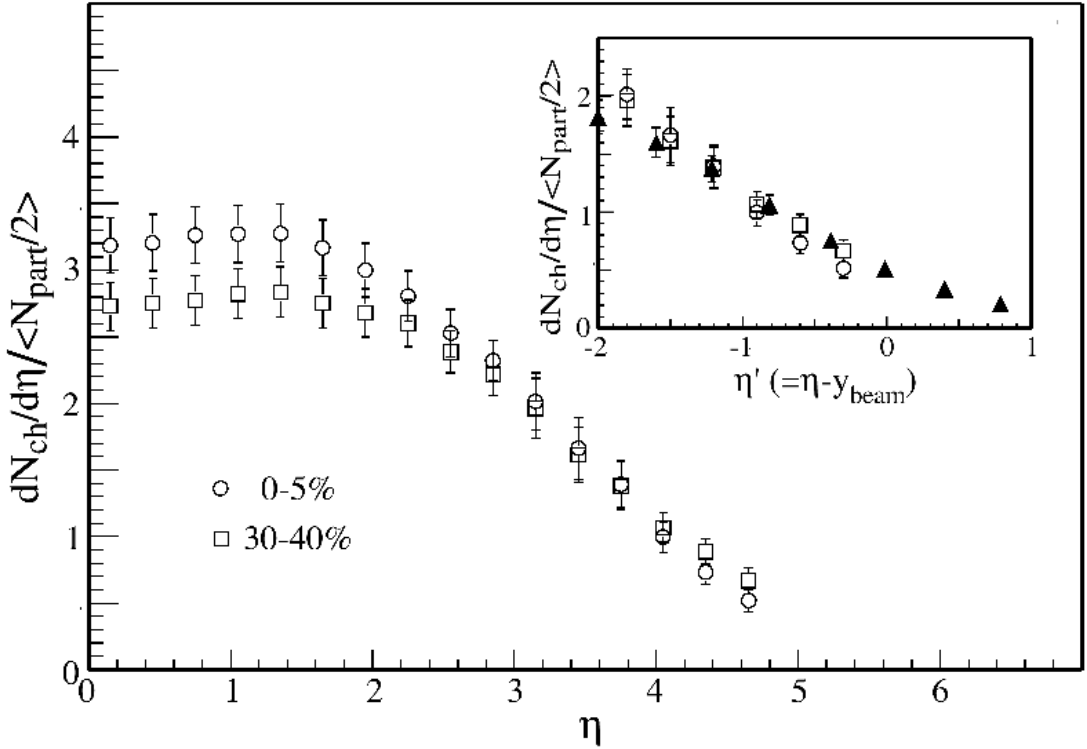,width=15.0cm}
\caption{
Charged particle densities normalized to the number of
participant pairs.  Total (statistical + systematic) uncertainties are shown,
not including the uncertainty in $<N_{part}>$ (see text).
The insert compares the 0-5\% central (open circles) and
30-40\% central (open squares) Au+Au results to the 9.4\% central 
Pb+Pb data (closed triangles)  of ref.~\cite{deines00}. 
For this comparison of the ``projectile'' regions for the two reactions, 
the data are plotted in terms of the  pseudorapidity shifted by the 
beam rapidity, as discussed in the text. For the insert, the
particle densities are normalized to the number of projectile 
participants, which is equal to the number of participant pairs for a 
symmetric reaction.
}
\label{dndeta_fragment}  
\end{figure}

Fig.~\ref{dndeta_fragment} shows the 
charged-particle multiplicity distributions scaled by half of the 
average number of participant nucleons 
for the 0-5\% and 30-40\% centrality cuts.
The average number of participants ($<N_{part}>$) corresponding to the
different centrality cuts was obtained using the HIJING code~\cite{wang91}, 
where the
collision geometry is based on the Glauber model~\cite{glauber}.  
The uncertainty in $<N_{part}>$ largely reflects the uncertainty
in the multiplicity measurement and is estimated as $<$0.5\% for 5\% 
centrality, increasing to 4\% at 50\% centrality.
For Fig. 3 (and Fig. 4), the 
$dN_{ch}/d\eta$ distributions have been symmetrized and
uncertainty weighted averages have been
developed for the different global detector subsystems.
It is found that near mid-rapidity, the charged-particle yield per 
participant pair is significantly greater for more central events as
compared to peripheral events.  However, at larger pseudorapidities, this
difference disappears. The mid-rapidity behavior is consistent with that 
observed by the
PHENIX~\cite{adcox01} and PHOBOS~\cite{back01} collaborations 
and attributed~\cite{adcox01} 
to an increased contribution of harder scattering processes for more
central collisions. 

The insert in Fig. 3 shows the Au+Au results for large
pseudorapidities in terms of the 
pseudorapidity shifted by the beam rapidity, thus allowing 
for a comparison with
SPS Pb+Pb data~\cite{deines00} within the ``limiting fragmentation'' 
picture~\cite{benecke69}.
This hypothesis states that at high energies the number of particles 
produced by the
``wounded projectile'' nucleons should be independent of the details of the
``target'', the ``projectile'', 
and the beam energy.   When shifted by the beam rapidity,
effectively translating to the beam's reference frame (assuming
similar values for the pseudorapidy and rapidity variables), 
the normalized yields near $\eta' (=\eta - y_{beam}) = 0$ should be system 
and energy
independent.  A ``limiting fragmentation'' behavior
has been observed  in
$pp$, $p\bar{p}$, $p$-emulsion, and $\pi$-emulsion data 
(see references 34-37 of ref.~\cite{deines00}).  It is shown in
ref.~\cite{deines00} that the hypothesis also works in comparing
Pb+Pb data at $\sqrt{s_{\small{NN}}}$=17.2 GeV with   
200 GeV/nucleon O+AgBr and S+AgBr results~\cite{dabrowska93}. 
The closed triangles in the Fig. 3 insert show the Pb+Pb results. Inspection
of this figure, where the current results at  $\sqrt{s_{\small{NN}}}$=130 GeV
follow the same general curve as found for the lower energy data,
suggests that the ``limiting fragmentation''  picture is successful over a
very large range of energies extending from SPS to RHIC 
($\sqrt{s_{\small{NN}}}$=17.2 GeV to 130 GeV). The absence of a centrality
dependence for the scaled yields near beam rapidity, 
as seen in Fig. 3, is also consistent with such a picture.

\begin{figure}
\epsfig{file=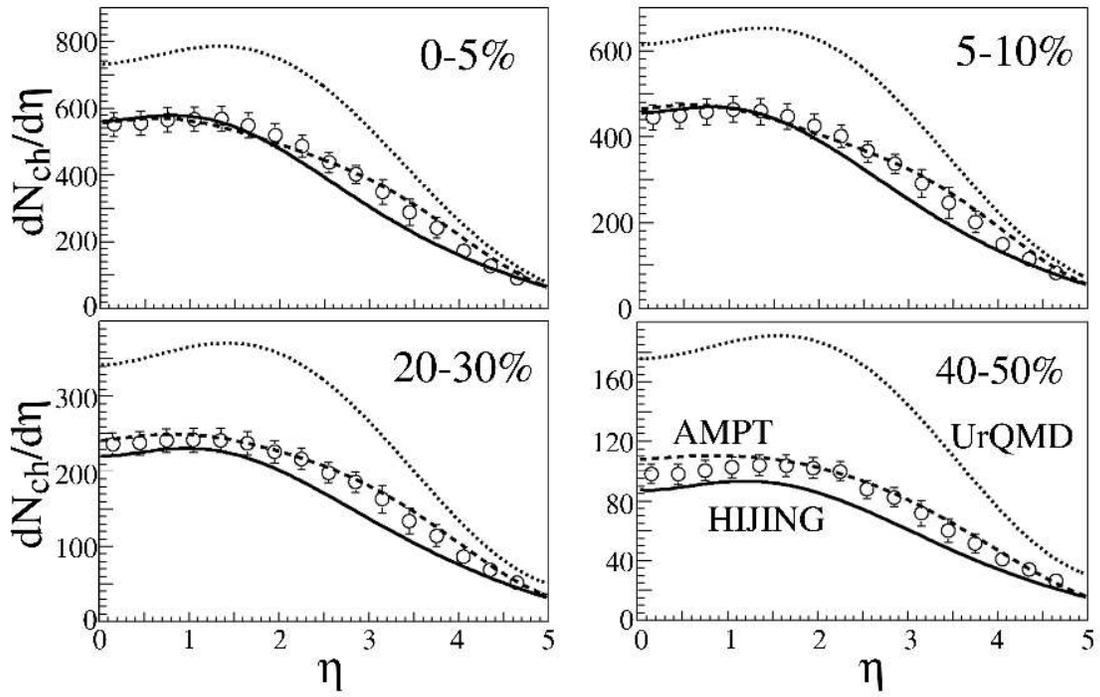,width=15.0cm}
\caption{
Distribution of  $dN_{ch}/d\eta$ for the indicated centrality
ranges.  Total uncertainties are indicated.
HIJING, AMPT, and UrQMD model calculations are shown. 
}
\label{dndeta_models}  
\end{figure}

In Fig.~\ref{dndeta_models} we compare HIJING~\cite{wang91},
UrQMD~\cite{urqmd},  and AMPT model~\cite{zhang01,lin01a,lin01b} 
calculations to the data for
different centrality
ranges.  HIJING is a model that places special
emphasis on  perturbative QCD processes leading to 
multiple minijet production from parton scatterings. The model, which
does not include final state rescattering except for schematic
jet quenching, is found to 
do a good job in describing the data at mid-rapidity, but
underestimates the widths of the multiplicity distributions at
all centralities.  The AMPT model uses HIJING to generate the
initial phase space of partons, but then extends the calculations to
model the parton-parton collisions and the final
hadronic interactions. By including these final state interactions,
the widths of the multiplicity distributions are found to agree well with
experiment, suggesting the importance of these final state interactions.
It is also interesting to note that the LUCIFER code~\cite{lucifer} 
gives similar results to AMPT but within a purely hadronic formulation. 
The UrQMD transport 
model accounts for 
hadronic interactions at low and intermediate energies in
terms of interactions between known hadrons and their resonances and
at higher energies in terms of excitations of color strings and their 
subsequent fragmentation.  
For the present RHIC data, the shape of the pseudorapidity distributions
are well reproduced, however with the particle production being way too large.
The current results appear to rule out the UrQMD model as a description of the 
RHIC data.
    
The BRAHMS experiment has measured pseudorapidity densities of 
charged particles from Au+Au collisions at $\sqrt{s_{NN}}$=130~GeV 
over a large range of pseudorapidity as a 
function of collision centrality using several independent detector 
systems and methods. 
We observe an enhancement of particle production for
central collisions at mid-rapidity, suggesting a breakdown in the simple
scaling by the number of participant nucleons seen at larger pseudorapidities.
Comparisons of the Au+Au fragmentation region data with SPS energy Pb+Pb
results suggest that a ``limiting fragmentation'' behavior 
is seen for nucleus-nucleus
collisions and is, in fact, already reached at the lower energy.
With the inclusion of final hadronic interactions, parton scattering
models are found to give a good description of the observed behaviors.

\ack
The BRAHMS collaboration wishes to thank the RHIC team for their
efforts leading to the successful startup of the collider and
for the support they have given to the experiment. 
This work was supported by the 
Division of Nuclear Physics
of the Office of Science of the U.S. Department of Energy under
contracts DE-AC02-98-CH10886, DE-FG03-93-ER40773, DE-FG03-96-ER40981, and
DE-FG02-99-ER41121, the Danish Natural Science Research Council, the 
Research Council of Norway, the Jagiellonian University Grants, 
the Korea Research Foundation Grant, and the Romanian Ministry of 
Education and Research (5003/1999,6077/2000).

\newpage
\begin{table}[h!]
\caption{\label{TABLE}\textit{\sl Charged particle densities in
$dN_{ch}/d\eta$ as a function of
centrality and pseudorapidity. Total uncertainties, dominated by
the systematics,  are indicated. The average
number of participants $<N_{part}>$ is given for each centrality
class based on HIJING model calculations. The last column gives the
integral charged particle multiplicity within the pseudorapidity
range $-4.7 \le \eta \le 4.7$.}}
\vspace{0.5cm}
\begin{tabular}{|c|c|c|c|c|c|c|}
\hline
Centrality &
$<N_{part}>$&
$\eta = 0$ &
$\eta =1.5$ &
$\eta = 3.0$ &
$\eta = 4.5$ &
$N_{ch}$
\\
\hline
0-5\%   & 352 & 553$\pm$36 & 554$\pm$37 & 372$\pm$37 & 107$\pm$15 & 3860$\pm$300 \\
\hline
5-10\%  & 299 & 447$\pm$29 & 454$\pm$31 & 312$\pm$36 &  94$\pm$13 & 3180$\pm$250 \\
\hline
10-20\% & 235 & 345$\pm$23 & 348$\pm$25 & 243$\pm$27 &  79$\pm$10 & 2470$\pm$190 \\
\hline
20-30\% & 165 & 237$\pm$16 & 239$\pm$16 & 172$\pm$18 &  59$\pm$8  & 1720$\pm$130 \\
\hline
30-40\% & 114 & 156$\pm$11 & 159$\pm$11 & 117$\pm$13 &  43$\pm$6  & 1160$\pm$90  \\
\hline
40-50\% &  75 &  98$\pm$7   & 104$\pm$7  & 77$\pm$9   &  30$\pm$4 &  750$\pm$60 \\
\hline
\end{tabular}
\end{table}

\newpage
\newpage
\newpage
\newpage

\end{document}